\documentclass{article}
\usepackage{amssymb}
\usepackage{amsfonts}
\usepackage{amsmath}

\input{tcilatex}

\begin{document}

\title{Nonextensive random-matrix theory based on Kaniadakis entropy }
\author{A.Y. Abul-Magd \\
Department of Mathematics, Faculty of Science,\\
Zagazig University, Zagazig, Egypt}
\maketitle

\begin{abstract}
The joint eigenvalue distributions of random-matrix ensembles are derived by
applying the principle maximum entropy to the R\'{e}nyi, Abe and Kaniadakis
entropies. While the R\'{e}nyi entropy produces essentially the same
matrix-element distributions as the previously obtained expression by using
the Tsallis entropy, and the Abe entropy does not lead to a closed form
expression, the Kaniadakis entropy leads to a new generalized form of the
Wigner surmise that describes a transition of the spacing distribution from
chaos to order. This expression is compared with the corresponding
expression obtained by assuming Tsallis' entropy as well as the results of a
previous numerical experiment.
\end{abstract}

\section{Introduction}

In 1957 Jaynes \cite{jaynes} proposed a rule, based on information theory,
to provide a constructive criterion for setting up probability distributions
on the basis of partial knowledge. This leads to a type of statistical
inference which is called the maximum-entropy principle (MaxEnt). It is the
least biased estimate possible on the given information. Jaynes showed in
particular how his rule, when applied to statistical mechanics, leads to the
usual Gibbs' canonical distribution. The core of the MaxEnt method resides
in interpreting entropy, through the Shannon axioms, as a measure of the
\textquotedblleft amount of uncertainty\textquotedblright\ or of the
\textquotedblleft amount of information that is missing\textquotedblright\
in a probability distribution. This was an important step forward because it
extended the applicability of the notion of entropy far beyond its original
roots in thermodynamics. In this paper we consider the application of MaxEnt
to the random-matrix theory (RMT), which is often used to describe quantum
systems whose classical counterparts have chaotic dynamics \cite{mehta,guhr}%
. This is the statistical theory of random matrices $H$ whose entries
fluctuate as independent Gaussian random numbers. Dyson \cite{dyson} showed
that there are three generic ensembles of random matrices, defined in terms
of the symmetry properties of the Hamiltonian. The most popular of these is
the Gaussian orthogonal ensemble (GOE) which successfully describes many
time-reversal-invariant quantum systems whose classical counterparts have
chaotic dynamics. Balian \cite{balian} derived the weight functions $P(H)$
for the random-matrix ensembles from MaxEnt postulating the existence of a
second moment of the Hamiltonian. He applied the conventional Shannon
definition for the entropy to ensembles of random matrices as%
\begin{equation}
S_{\text{Sch}}=-\int dHP(H)\ln P(H)
\end{equation}%
and maximized it under the constraints of normalization of $P(H)$\ and fixed
mean value of Tr$\left( H^{T}H\right) $, where the superscript $T$ denotes
the transpose. The latter constraint ensures basis independence, which is a
property of the trace of a matrix. Then, the distribution $P(H)$ is
determined from the extremum of the functional%
\begin{equation}
F_{\text{Sch}}=S_{\text{Sch}}-\xi \int dH~P(H)-\eta \int dH~P(H)\text{Tr}%
\left( H^{T}H\right) ,
\end{equation}%
where $\xi $ and $\eta $ are Lagrange multipliers. Its maximum is obtained
equating its functional derivative to zero. He obtained%
\begin{equation}
P(H)\varpropto \exp \left[ -\eta \text{Tr}\left( H^{T}H\right) \right]
\end{equation}%
which is a Gaussian distribution with inverse variance $1/2\eta $. Most of
the interesting results are obtained for $N\times N$ matrices in the limit
of $N$ $\rightarrow \infty $. A complete discussion of the level
correlations for large matrices is a difficult task. Analytical results have
long ago been obtained for the case of $N=2$ \cite{porter}. The distribution
of nearest-neighbor-spacings (NNS) of levels for GOE of 2 dimensional
matrices,%
\begin{equation}
P(s)=\left( \pi /2\right) s\exp \left( -\pi s^{2}/4\right) ,
\end{equation}%
is known as Wigner's surmise. Here $s$ is scaled so that the mean spacing
equal one. Analogous expression are obtained for the Gaussian unitary and
symplectic ensembles. To demonstrate the accuracy of the Wigner surmise, we
expand this distribution in powers of $s$ to obtain%
\begin{equation}
P(s)=\frac{32}{\pi ^{2}}s^{2}\left( 1-\frac{4}{\pi }s^{2}+\cdots \right)
\cong 3.242s^{2}-4.128s^{4}+\cdots ,
\end{equation}%
while the power-series expansion of the corresponding exact distribution 
\cite{mehta} yields%
\begin{equation}
P_{\text{exact}}(s)=\frac{\pi ^{2}}{3}s^{2}-\frac{2\pi ^{4}}{45}s^{4}+\cdots
\cong 3.290s^{2}-4.329s^{4}+\cdots .
\end{equation}%
The Wigner surmise has been successfully applied to the NNS distributions
for numerous chaotic systems.

For most systems, the phase space is partitioned into regular and chaotic
domains. These systems are known as mixed systems. Attempts to generalize
RMT to describe such mixed systems are numerous; for a review please see 
\cite{guhr}. MaxEnt can contribute to the generalization of RMT by either
introducing additional constraints or modifying the entropy. An example of
the first approach is the work of Hussein and Pato \cite{hussein}, who use
MaxEnt to construct "deformed" random-matrix ensembles by imposing different
constraints for the diagonal and off-diagonal elements.

The second approach to include mixed systems in RMT is to apply MaxEnt to
entropic measures other than Schannon's. Recently, the Tsallis statistical
mechanics \cite{Ts1,Ts2} has been applied to include systems with mixed
regular-chaotic dynamics in a nonextensive generalization of RMT \cite%
{evans,toscano,bertuola,nobre,abul1,abul2}. Different entropic forms have
been introduced, like Tsallis', by appropriately deforming the logarithm in
the expression for the Shannon entropy \cite{naudts}. Among the many
generalizations, one can find the entropies considered by R\'{e}nyi \cite%
{renyi}, by Abe \cite{abe,abe1}, by Landsberg and Vedral \cite{landsberg},
and by Kaniadakis ($\kappa $-entropy) \cite%
{kaniadakis,kaniadakis1,kaniadakis2}. The present paper is a natural
continuation of the previous work on nonextensive RMT, by considering other
forms of generalized entropy.

\section{Nonextensive RMT}

In this section, we review the main results of non-extensive RMT based on
Tsallis entropy, and show that replacing it with other generalized entropies
does not always produce new results. The matrix-element distribution $P(H)$
is obtained by extremizing Tsallis' $q$-entropy 
\begin{equation}
S_{\text{Ts}}\left[ q,P(q,H)\right] =\left. \left( 1-\int dH\left[ P_{\text{%
Ts}}(q,H)\right] ^{q}\right) \right/ (q-1).
\end{equation}%
rather than Shannon's, but again subject to the same constraints of
normalization and existence of the expectation value of Tr$\left(
H^{T}H\right) $. This can be done by replacing $S_{\text{Sch}}$ in Eq. (2)
by $S_{\text{Ts}}$ and then equating the functional derivative in the
resulting expression to zero. The extremization then yields%
\begin{equation}
P_{\text{Ts}}(q,H)=Z_{q}^{-1}\left[ 1+(q-1)\eta _{q}\text{Tr}\left(
H^{T}H\right) \right] ^{-\frac{1}{q-1}},
\end{equation}%
where $Z_{q}$ and $\eta _{q}$ are expressed in terms of the Lagrange
multipliers. The case of 2-dimensional ensembles belonging to the three
symmetry classes is considered in Ref. \cite{abul1}. For ensembles with
orthogonal symmetry, the resulting NNS distribution is given by%
\begin{equation}
P_{\text{Ts}}(s)=\left( \frac{1}{q-1}-\frac{3}{2}\right) \eta _{q}s\left[ 1+%
\frac{1}{2}(q-1)\eta _{q}s^{2}\right] ^{-1/(q-1)+1/2},
\end{equation}%
for $q>1$, where%
\begin{equation}
\eta _{q}=\frac{\pi }{2}\left( \frac{\Gamma \left[ (1/(q-1)-2\right] }{%
\Gamma \left[ 1/(q-1)-3/2\right] }\right) ^{2},
\end{equation}%
and $\Gamma \lbrack x]$ is Euler's gamma function \cite{gradshteyn}. The
extension of these results to the case on ensembles with arbitrarily large $%
N $ amounts to replacing $q$ in Eq. (9) by another $N$-dependent parameter 
\cite{abul2}.

It is well known that the Shannon entropy of the couple of two independent
random variables is the sum of their respective Shannon entropies. We have
firstly attempted to extremize R\'{e}nyi's entropy for the matrix-element
probability density function $P_{\text{Rn}}(H),$ which is defined by%
\begin{equation}
S_{\text{Rn}}\left[ \alpha ,P_{\text{Rn}}(H)\right] =\frac{1}{1-\alpha }\ln
\left( \int dH\left[ P_{\text{Rn}}(H)\right] ^{\alpha }\right) ,
\end{equation}%
where the index $\alpha ~$varies between 1 and 2.The obtained distribution
NNS is very similar to that in Eq. (9), which is obtained by the
maximization of Tsallis' entropy,\ if we replace $1-q$ by $\alpha -1$. In
other words, mere fittings of observed data by the NNS distributions in Eq.
(9) do not tell us anything about which of Tsallis' or R\'{e}nyi's entropy
is the underlying physical quantity. This result agrees with the known fact 
\cite{lenzi} that the maximum entropy principle in combination with R\'{e}%
nyi's entropy reproduces the equilibrium probability distributions of
Tsallis' non-extensive thermostatistics. In fact,\ the R\'{e}nyi entropy is,
in all cases, a monotonically increasing function of the Tsallis entropy%
\begin{equation}
P_{\text{Rn}}=\left. \frac{\ln \left[ 1+(1-q)P_{\text{Ts}}\right] }{1-q}%
\right\vert _{q=\alpha }.
\end{equation}%
Therefore the probability density function that extremizes one of these
entropies automatically extremizes the other.

Abe's entropy \cite{abe} is another interesting example of non-extensive
entropies. It applies a $q$-calculus, which is invariant under the exchange $%
q\rightarrow q^{-1}$. \ This is a symmetry which plays a central role in the
physical context of quantum groups \cite{chaikin}. This entropy has been
introduced in \cite{abe,abe1} and has been applied there to generalized
statistical-mechanical study of $q$-deformed oscillators. Abe's entropy,
generated for the distribution function of a random-matrix ensemble, is
defined by%
\begin{equation}
S_{\text{Abe}}\left[ q,P(q,H)\right] =\left. -\int dH\left( \left[ P_{\text{%
Abe}}(q,H)\right] ^{q}-\left[ P_{\text{Abe}}(q,H)\right] ^{-q}\right)
\right/ (q-q^{-1}),
\end{equation}%
and also reduces to Schannon's entropy in the $q\rightarrow 1$\ limit. Due
to the $q\rightarrow q^{-1}$ symmetry the range of $q$ may be restricted to
0 \TEXTsymbol{<} $q$ \TEXTsymbol{<} 1. Now, replacing $S_{\text{Sch}}$ in
Eq. (2) by $S_{\text{Abe}}$ and then equating the functional derivative in
the resulting expression to zero, one obtains after some reorganizations%
\begin{equation}
-\frac{qP_{\text{Abe}}^{q-1}-q^{-1}P_{\text{Abe}}^{q^{-1}-1}}{q-q^{-1}}-\xi
_{\text{Abe}}-\eta _{\text{Abe}}\text{Tr}\left( H^{T}H\right) =0.
\end{equation}%
Unfortunately, it is not possible to solve this equation with respect to $P_{%
\text{Abe}}$ analytically. The analogous equation is solved numerically in
Ref. \cite{abe}. When $q$ = 1, one obtains the standard GOE distribution,
whereas for 0 \TEXTsymbol{<} $q$ \TEXTsymbol{<} 1, the distribution comes to
exhibit the power-law behavior similar to similar to the distributions
obtained by extremizing the\ Tsallis and Renyi entropies. Indeed, in the
asymptotic region where $P_{\text{Abe}}\ll 1$, one neglects the term in
which $P_{\text{Abe}}$ is raised to a positive power and obtains%
\begin{equation}
P_{\text{Abe}}(q,H)~\sim \left[ \xi _{\text{Abe}}+\eta _{\text{Abe}}\text{Tr}%
\left( H^{T}H\right) \right] ^{-1/(1-q)}.
\end{equation}

\section{Kaniadakis' entropy}

In this section, we consider a possible generalization of RMT based on an
extremization of Kaniadakis' $\kappa $-entropy \cite%
{kaniadakis,kaniadakis1,kaniadakis2}. This entropy shares the same symmetry
group of the relativistic momentum transformation and has applications in
cosmic-ray and plasma physics. For the matrix-element probability
distribution function, it reads%
\begin{equation}
S_{\text{K}}\left[ \kappa ,P_{\text{K}}(\kappa ,H)\right] =-\frac{1}{2\kappa 
}\int dH\left( \frac{\alpha ^{\kappa }}{1+\kappa }\left[ P_{\text{K}}(\kappa
,H)\right] ^{1+\kappa }-\frac{\alpha ^{-\kappa }}{1-\kappa }\left[ P_{\text{K%
}}(\kappa ,H)\right] ^{1-\kappa }\right)
\end{equation}%
with $\kappa $ a parameter with value between 0 and 1; the case of $\kappa
=0 $, corresponds to the Schannon entropy. Here, $\alpha $ is a real
positive parameter. Kaniadakis has considered two choices of $\alpha $,
namely $\alpha =1$ and $\alpha =Z$, where $Z$ is the generalized partition
function. We here adopt the second choice. The matrix-element distribution $%
P_{\text{K}}(\kappa ,H)$ is obtained by extremizing the functional%
\begin{equation}
F_{\text{K}}=S_{\text{K}}-\eta _{\text{K}}\int dH~P_{\text{K}}(\kappa ,H)%
\text{Tr}\left( H^{T}H\right) ,
\end{equation}%
where $\eta _{\text{K}}$ is a Lagrange multiplier. One arrives to the
following distribution%
\begin{equation}
P_{\text{K}}(\kappa ,H)=\frac{1}{Z}\exp _{\left\{ \kappa \right\} }\left[
-\eta _{\text{K}}\text{Tr}\left( H^{T}H\right) \right] ,
\end{equation}%
where%
\begin{equation}
Z=\int dH~\exp _{\left\{ \kappa \right\} }\left[ -\eta _{\text{K}}\text{Tr}%
\left( H^{T}H\right) \right] .
\end{equation}%
Here $\exp _{\left\{ \kappa \right\} }\left[ x\right] $ is the $\kappa $%
-deformed exponential \cite{kaniadakis} which is defined by 
\begin{equation}
\exp _{\left\{ \kappa \right\} }\left[ x\right] =\left( \sqrt{1+\kappa
^{2}x^{2}}+x\right) ^{1/\kappa }=\exp \left( \frac{1}{\kappa }\text{arcsinh }%
\kappa x\right) .
\end{equation}%
and has the properties $\exp _{\left\{ 0\right\} }\left[ x\right] =\exp (x),$
$\exp _{\left\{ \kappa \right\} }\left[ 0\right] =1$ and $\exp _{\left\{
-\kappa \right\} }\left[ x\right] =\exp _{\left\{ \kappa \right\} }\left[ x%
\right] ,$ and obeys the scaling law $\exp _{\left\{ \kappa \right\} }\left[
\lambda x\right] =(\exp _{\left\{ \kappa /\lambda \right\} }\left[ x\right]
)^{\lambda }$.The asymptotic behavior of the $\kappa $-exponential is $\exp
_{\left\{ \kappa \right\} }\left[ x\right] \underset{x\rightarrow \pm \infty 
}{\sim }$ $\left\vert 2\kappa x\right\vert ^{\pm 1/\left\vert \kappa
\right\vert }.$

To calculate $Z$, we note that Tr$\left( H^{T}H\right)
=\sum_{i=1}^{N}H_{ii}^{2}+2\sum_{i>j}H_{ij}^{2}$. We introduce the new
coordinates $\mathbf{r}=\left\{ r_{1},\cdots ,r_{n}\right\} $, where $%
n=N(N+1)/2$, and $r_{i}^{2}$ stand for the square of the diagonal elements
or twice the square of the non-diagonal elements, respectively. With these
new variables, Eq. (19) becomes%
\begin{multline}
Z=2^{-N(N-1)/4}\int d^{n}\mathbf{r}\exp \left( -\frac{1}{\kappa }\text{%
arcsinh }\kappa \eta _{\text{K}}r^{2}\right) \\
=2^{-N(N-1)/4}\left( \frac{\pi }{2\kappa \eta _{\text{K}}}\right) ^{n/2}%
\frac{\Gamma \left( \frac{1}{2\kappa }-\frac{n}{4}\right) }{\left( 1+n\kappa
/2\right) \Gamma \left( \frac{1}{2\kappa }+\frac{n}{4}\right) }
\end{multline}%
for $\kappa <2/n$. Here we use the result obtained by Kaniadakis in
evaluating an analogous integral that appears in his treatment of Brownian
particles (Eq. (71) of Ref. \cite{kaniadakis}).

We now calculate the joint probability density for the eigenvalues of the
Hamiltonian $H$. With $H=U^{-1}EU$, where $U$\ is the global unitary group,
we introduce the elements of the diagonal matrix of eigenvalues $E=$ diag$%
(E_{1},\cdots ,E_{N})$ of the eigenvalues and the independent elements of $U$
as new variables. Then the volume element $dH$ has the form 
\begin{equation}
dH=\left\vert \Delta _{N}\left( E\right) \right\vert dEd\mu (U),
\end{equation}%
where $\Delta _{N}\left( E\right) =\prod_{n>m}(E_{n}-E_{m})$ is the
Vandermonde determinant and $d\mu (U)$ the invariant Haar measure of the
unitary group \cite{mehta,guhr}. The probability density $P_{\text{K}%
}(\kappa ,H)$ depends on $H$ through Tr$\left( H^{\dagger }H\right) $\ and
is therefore invariant under arbitrary rotations in the matrix space. \
Integrating over the "angular variables" $U$ yields the joint probability
density of eigenvalues in the form%
\begin{equation}
P_{\text{K}}(\kappa ;E_{1},...,E_{N})=C\prod_{n>m}(E_{n}-E_{m})\exp
_{\left\{ \kappa \right\} }\left[ -\eta _{\text{K}}\sum_{i=1}^{N}E_{i}^{2}%
\right] .
\end{equation}%
where $C$ is a normalization constant.

In order to obtain a generalization of the Wigner surmise, we consider the
case of two-dimensional random-matrix ensemble where $N=2$ and $n=3.$ In
this case, Eq. (23) reads%
\begin{equation}
P_{\text{K}}(\kappa ;\varepsilon ,s)=\frac{2\left( 1+3\kappa /4\right) }{%
B\left( \frac{1}{2\kappa }-\frac{3}{4},\frac{3}{2}\right) }\left( \kappa
\eta _{\text{K}}\right) ^{3/2}s\exp _{\left\{ \kappa \right\} }\left[ -\eta
_{\text{K}}\left( 2\varepsilon ^{2}+\frac{1}{2}s^{2}\right) \right] ,
\end{equation}%
where $\varepsilon =(E_{1}+E_{2})/2,~s=\left\vert E_{1}-E_{2}\right\vert $,
and $B(a,b)=\Gamma (a)\Gamma (b)/\Gamma (a+b)$ is the Beta function \cite%
{gradshteyn}. The NNS distribution is obtained by integrating (24) over $%
\varepsilon $ from $-\infty $ to $\infty $. This can be done by changing the
variable $\varepsilon $ into $x=\exp [-\frac{1}{\kappa }$arcsinh$(\kappa
\eta _{\text{K}}s^{2}/2)]$, integrating by parts, and then replacing the
variable $x$ by another new variable, $y=\exp (\kappa x)$. The resulting
integral can be solved by using the following identity \cite{gradshteyn}%
\begin{multline}
\int_{u}^{\infty }y^{-\lambda }(y+\beta )^{\nu }\left( y-u\right) ^{\mu
-1}dy=u^{\mu +\nu -\lambda }B\left( \lambda -\mu -\nu ,\mu \right) \\
~{}_{2}F_{1}\left( -\nu ,\lambda -\mu -\nu \,;\lambda -\nu ;-\frac{\beta }{u}%
\right) ,
\end{multline}%
for $\left\vert \beta /u\right\vert \,<1$ and $0<\mu <\lambda -\nu $,\ where 
$_{2}F_{1}(\nu ,\mu \,;\lambda ;x)$ is the hypergeometric function. Thus,
after straightforward calculations we can express the NNS\ distribution as 
\begin{multline}
P_{\text{K}}(s)=-2\left( 1+\frac{3}{4}\kappa \right) \eta _{\text{K}%
}se^{\left( 1/2-1/\kappa \right) \text{arcsinh}(\kappa \eta _{\text{K}%
}s^{2}/2)}\frac{B\left( \frac{1}{\kappa }-\frac{1}{2},\frac{3}{2}\right) }{%
B\left( \frac{1}{2\kappa }-\frac{3}{4},\frac{3}{2}\right) }{} \\
_{2}F_{1}\left( -\frac{1}{2},\frac{1}{\kappa }-\frac{1}{2}\,;\frac{1}{\kappa 
}+1;-e^{-2\text{arcsinh}(\kappa \eta _{\text{K}}s^{2}/2)}\right) .
\end{multline}%
The condition of unit mean spacing defines the quantity $\eta _{\text{K}}$\
as%
\begin{equation}
\eta _{\text{K}}=\left[ \frac{\pi k^{3/2}\left( 1+\frac{3}{4}\kappa \right) 
}{\left( 1-\kappa ^{2}\right) B\left( \frac{1}{2\kappa }-\frac{3}{4},\frac{3%
}{2}\right) }\right] ^{2}.
\end{equation}%
We note that the function $B\left( \frac{1}{\kappa }-\frac{1}{2},\frac{3}{2}%
\right) $ diverges at $\kappa =\kappa _{c}=1/2$, which serves as an upper
limit for the range of variation of $\kappa $. We also note that the mean
square spacing diverges at $\kappa =\kappa _{\infty }=2/5$.

We now compare the spacing distributions $P_{\text{K}}(s)$ and $P_{\text{Ts}%
}(s).$ Each of the the two distributions coincides with the Wigner
distribution for the smallest values of the entropic indices $\kappa =0$ and 
$q=1$, but essentially differs from the Poisson distribution and practically
behaves as a delta function $\delta (s)$ at the maximum allowed values of $%
\kappa =0.5$ and $q=1.5$. Therefore, neither $P_{\text{K}}(s)$ nor $P_{\text{%
Ts}}(s)$ can describe the initial stage of the transition from chaos to
integrability. In fact, it is hard to believe that nearly integrable systems
can be described by a base-independent random-matrix model like the one
considered in the present paper. Both distributions increase linearly with $%
s $ near the origin. The Kaniadakis distribution $P_{\text{K}}(s)$ decays at
large $s$ as $s^{2-2/\kappa }$. The NNS in Eq. (9) obtained by the Tsallis
statistics asymptotically behaves as $P_{\text{Ts}}(s)\sim s^{2-2/(q-1)}$.
Both distributions have similar asymptotic behavior when $\kappa =q-1$.
However, distributions $P_{\text{K}}(s)$ and $P_{\text{Ts}}(s)$ satisfying
this relation have different behavior at intermediate values of $s$. The
evolution of $P_{\text{K}}(s)$ as $\kappa $ varies from the GOE value of 0
to $\sim 0.5$ and $P_{\text{Ts}}(s)$ as $q$ varies from the GOE value of 1
to $\sim 1.5$ is demonstrated in Fig.1. The figure shows two sets of NNS
distributions, one set for each statistics, with entropic indices related by 
$\kappa =q-1$\ so that the corresponding distribution have similar
asymptotic behavior. While the peak of the Tsallis distribution steadily
moves towards the origin as $q$ increases, the peak of the Kaniadakis
distribution remains almost in the same position as $\kappa $ increases from
0 and only starts to move towards smaller $s$ when $\kappa $ exceeds a value
of $\sim 0.4$. We remind that the mean square spacing diverges in this
latter domain of $\kappa $.

\section{Comparison with numerical experiment}

This section offers a comparison of the generalized expressions of the
Wigner surmise derived from the Tsallis and Kaniadakis entropies with the
NNS obtained in the numerical experiment done by \.{Z}yczkowski and Ku\'{s} 
\cite{zyc}. This experiment imitates a one-parameter (denoted by $\delta $)
transition between an ensemble of diagonal matrices with independently and
uniformly distributed elements, corresponding to $\delta =0,$ and a circular
orthogonal ensemble for $\delta =1$. These authors were able to achieve
reliable statistics by constructing numerically 500 matrices of size $N=100$
for each considered value of the transition parameter $\delta $.
Diagonalizing these matrices yielded NNS distributions consisting of 50 000
spacings for each value of $\delta $. Three of these distributions
intermediate between the Wigner and Poisson distributions, corresponding to $%
\delta =0.9,~0.5$ and 0.1, are reported in Ref. \cite{zyc}. These
distributions are compared in Fig. 1 with the spacing distributions in Eqs.
(9) and (26), which have been obtained by using the Tsallis and Kaniadakis
statistics, respectively. The figure suggests that both entropies yield
reasonable description of the stochastic transition although the agreement
with the numerical results is not complete. The slight difference in the
behavior of the two non-extensive NNS distributions is a positive aspects
since it is well known that the transition from order to chaos does not
necessarily proceed through the same passage. It remains an open question to
find which systems are better described by either entropy.

\section{Conclusion}

Tsallis' entropy has been considered by several authors as a starting point
for constructing a generalization of RMT. However, there are several other
generalized entropies. This paper derives random-matrix ensembles by
maximizing the generalized entropies proposed by R\'{e}nyi, Abe and
Kaniadakis under the constraints on normalization of the distribution and
the expectation value of Tr$\left( H^{T}H\right) $. The distribution
functions obtained for the R\'{e}nyi entropy are analogous to the
corresponding results, previously obtained assuming Tsallis' entropy, which
is not a surprise since these entropies are monotonic functions of each
other. The use of Abe's entropy did not lead to closed-form expressions for
the matrix-element distribution. On the other hand, we obtain a new
analytical formula for the distribution function in the matrix-element space
when the entropy is given by the Kaniadakis measure. This leads to a new
expression for the NNS distribution of the eigenvalues in the special case
of two-dimensions random-matrix ensembles. This special case is the one that
leads to the Wigner surmise when the entropy is given by the Schannon
measure. The high accuracy of Wigner's distribution describing chaotic
systems justifies the use of the two-dimensional ensembles for obtaining NNS
distributions using other entropies. As in the case of Tsallis' entropy, the
NNS distribution obtained here for Kaniadakis' entropy has an intermediate
shape between the Wigner distribution and the Poissonian that describes
generically integrable systems. However, neither of these distributions
reaches the Poissonian form for any value of the entropic index. It is
generally believed that the route from integrability to chaos is not unique.
Therefore, we hope that the NNS distribution obtained in this paper as well
as the one, previously obtained with the Tsallis statistics, are good
candidates for modelling systems with mixed regular-chaotic dynamics at
least when they are not far from the state of chaos. We have tested these
expressions by comparing their predictions with the NNS distributions
obtained in a numerical experiment by \.{Z}yczkowski and Ku\'{s}.

\bigskip

{\LARGE Figure caption}

\bigskip

FIG. 1. NNS distributions obtained by using the Tsallis (lower panel) and
Kaniadakis (upper panel) statistics. The entropic indices used, being $q=1$
(GOE)$,1.1,1.2,1.3,1.4,1.45,1.47,1.49$ in the case of the Tsallis statistics
and $\kappa =0$ (GOE)$,0.1,0.2,0.3,0.4,0.45,0.47,0.49~$in the Kaniadakis',
are chosen so that the corresponding curves in the two statistics have the
same asymptotic behavior.

FIG. 2. NNS distributions obtained in a numerical simulation \cite{zyc} of
the transition from the GOE statistics to the Poissonian (histograms)
compared to NNS distributions obtained by using the Tsallis (dashed lines)
and Kaniadakis (solid lines) statistics.

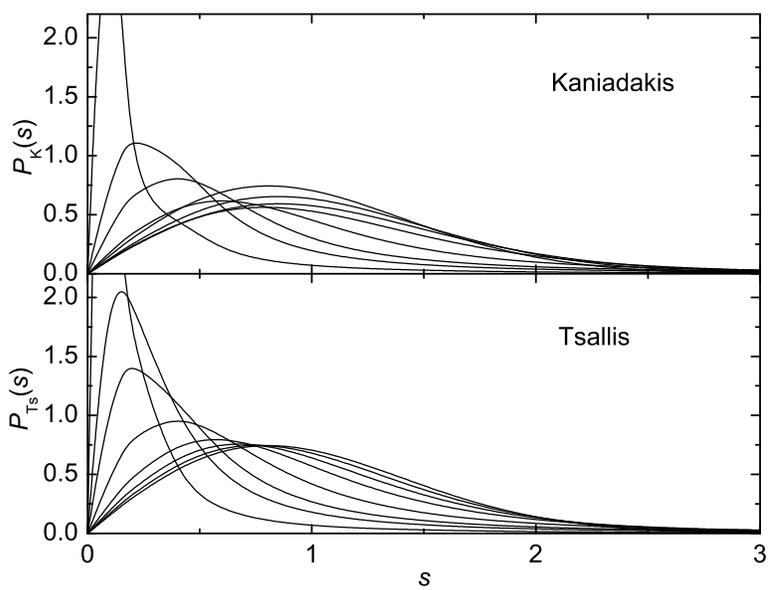
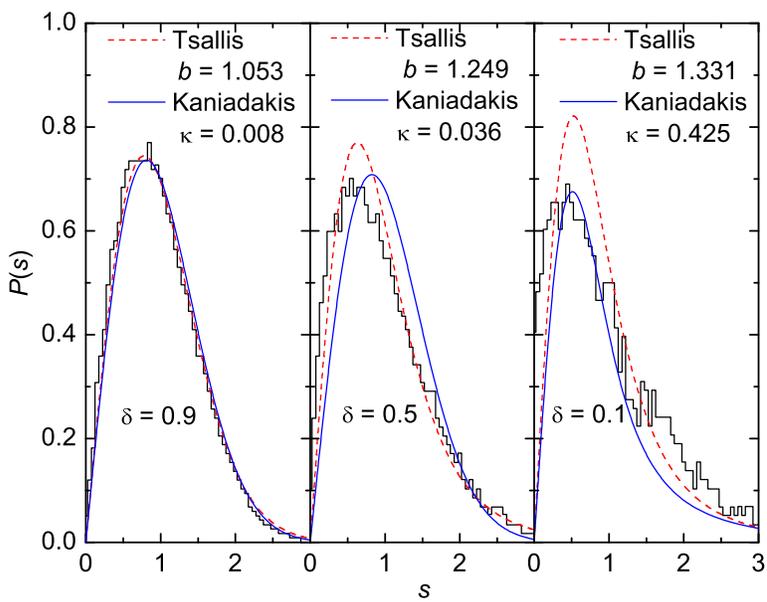
\end{document}